\documentclass[runningheads]{llncs}

\usepackage[T1]{fontenc}
\usepackage{graphicx}
\usepackage{subcaption}
\usepackage{calc}
\usepackage{amssymb}
\usepackage{amstext}
\usepackage{amsmath}
\usepackage{multicol}
\usepackage{algorithm2e}
\usepackage{xcolor}
\usepackage{svg}
\usepackage[bottom]{footmisc}

\usepackage{listings}

\usepackage{enumitem}
\usepackage{tikz}
\usetikzlibrary{shapes.geometric, arrows.meta, positioning, calc, backgrounds, fit}

\begin{document}

\title{Gray-Box Poisoning of Continuous Malware Ingestion Pipelines}

 \author{Jan Dolejš\orcidID{0009-0003-4396-3026} \and
 Martin Jureček\orcidID{0000-0002-6546-8953} \and
 Róbert Lórencz\orcidID{0000-0001-5444-8511}}
 
 \authorrunning{J. Dolejš et al.}

 \institute{Faculty of Information Technology, Czech Technical University in Prague, Thákurova 9, Prague, 16000, Czech Republic \\
 \email{\{dolejj13, martin.jurecek, robert.lorencz\}@fit.cvut.cz}}

\maketitle

\begin{abstract}
Modern malware detection pipelines rely on continuous data ingestion and machine 
learning to counter the high volume of novel threats. This work investigates 
a realistic gray-box poisoning threat model targeting these pipelines. 
Using the \texttt{secml\_malware} framework, we generate problem-space 
adversarial binaries through functionality-preserving manipulations, 
specifically Import Address Table (IAT) and section injections. 
We evaluate the impact of these poisoned samples when ingested into 
a defender's training set for a LightGBM malware detection model. 
Our empirical results demonstrate that subtle IAT-based perturbations 
enable compact poisoning samples that significantly degrade detection recall. 
These findings illustrate the inherent challenge of developing 
low-visibility adversarial perturbations that maintain high poisoning 
efficacy within continuous learning systems. We further evaluate a defense 
mechanism based on a homogeneous ensemble, which successfully 
identifies and filters up to 95.6\% of poisoning attempts while maintaining 
a high retention rate for legitimate data. These findings emphasize 
the necessity of robust pre-ingestion validation in production pipelines.

\end{abstract}

\section{Introduction}
\label{sec:introduction}

The rapid evolution of cyber threats necessitates highly scalable and
automated malware detection systems. With estimates suggesting approximately
560,000 new malware samples identified daily \cite{huntress}, security
vendors have shifted from manual analysis to machine learning (ML)
architectures integrated into continuous data ingestion pipelines
\cite{ucci2019survey}. These pipelines function as critical software
infrastructure, enabling the automated collection, feature extraction,
and retraining of models to maintain detection efficacy against emerging
variants.

However, the automated nature of these MLOps (Machine Learning Operations)
pipelines introduces a significant vulnerability in the data supply chain.
While evasion attacks \cite{adversarialdemetrio,gammademetrio,mabMW,pesidious}
target the model at inference time, a more insidious and persistent threat
is data poisoning. In a poisoning attack, an adversary injects malicious
samples into the ingestion pipeline, effectively "corrupting" the training
set to induce erroneous model behavior in future versions. From a software
engineering perspective, this represents a failure in the reliability and
integrity of the automated evolution process, where the system's learning
logic is subverted to create intentional backdoors or degrade overall
classification performance \cite{surveyaryal}.

In this paper, we investigate the robustness of continuous malware
ingestion pipelines against problem-space poisoning attacks. Unlike
feature-space attacks that manipulate abstract vectors, we focus on
functionality-preserving manipulations of actual PE binaries using the
\texttt{secml\_malware} framework \cite{secmlmalware}. We simulate a
realistic engineering scenario where an adversary leverages subtle
perturbations, such as Import Address Table (IAT) and section injections,
to bypass initial scrutiny and poison a LightGBM-based detection system.

To address this threat, we propose and evaluate a defense architecture
based on a homogeneous ensemble of oracles. This ensemble acts as a
quality-assurance middleware, filtering suspicious samples before they
reach the primary training distribution. Our evaluation focuses on the
trade-offs between ingestion throughput and security, providing insights
into the design of resilient automated detection systems.

The paper is structured as follows: Section~\ref{sec:poisoning} formulates
the threat model in the context of automated pipelines;
Section~\ref{sec:dataset} describes the engineering of the hybrid dataset;
Section~\ref{sec:experiments} presents empirical results on attack efficacy
and defense performance; Section~\ref{sec:limitations} discusses
architectural constraints, limitations and future work;
and Section~\ref{sec:conclusion} concludes.

\section{Poisoning Attacks}
\label{sec:poisoning}
Data poisoning compromises the integrity of machine learning models
by injecting adversarial samples into training sets \cite{nistAI}.
In malware detection, this typically involves inserting functionally
malicious binaries mislabeled as benign. By manipulating the feature
distribution, an attacker induces the model to learn erroneous correlations,
mapping malicious features to benign labels.

\subsection{Inverse-Feature Mapping and Evasion}
Many techniques focus on feature-space poisoning
\cite{featurepoisoningchen,poisoningseveri} or label-flipping
\cite{labelflippingsasaki,labelflippingaryal}. However, direct feature-space
modifications often result in abstract vectors lacking valid
executable counterparts, a challenge known as the inverse-feature
mapping problem \cite{poisoningpierazzi}. Realism requires attacks
in the problem space via actual binary perturbations.

We use \texttt{secml\_malware} \cite{secmlmalware} to generate viable
binaries using two approaches: \textbf{Gradient-Based Manipulations},
which identify optimal byte-level modifications \cite{adversarialdemetrio},
and \textbf{Heuristic-Based Manipulations}, which target PE structures
via genetic algorithms \cite{gammademetrio}. Both maintain malicious
functionality while evading detection.

\subsection{Poisoning Threat Model}
\label{sec:threat_model}
We formalize the interaction between a \textit{Defender} and an \textit{Adversary} in a
gray-box setting. Let $\mathcal{X}$ denote the problem space of all possible
PE binaries. The Defender utilizes a feature extraction mapping $\phi_D$ and
a detection model $M_D$, while the Adversary employs a mapping $\phi_A$
and a proxy model $M_A$. We assume $\phi_A \approx \phi_D$ (gray-box assumption)
and denote the shared mapping as $\phi$.

The Defender maintains a private training set $D_T$, an ingestion source $D_I$,
an oracle subset $D_O$, and a holdout set $D_H$ (unknown to both the Adversary
and Defender during training). New instances $x \in D_I$
are processed by a decision function $\Gamma: \mathcal{X} \to \{0, 1, \perp\}$,
where $0, 1$ represent benign and malicious labels, and $\perp$ denotes an
unlabeled/rejected state. 

The Adversary generates poisoned samples $D_P$ by performing evasion against
$M_A$ using public data $D_{pub} \subset D_I$. They seek to minimize a loss
function $\mathcal{L}$ such that $M_A(\phi(x))$ is misclassified as benign.
The resulting poisoned training set is:
\begin{equation}
D_{T}' = D_{T} \cup \{ (x, \Gamma(x)) \mid x \in D_{I} \cup D_{P} \text{ s.t. } \Gamma(x) \neq \perp \}
\end{equation}

\vspace{-0.3cm}
\section{Dataset Construction and Partitioning}
\label{sec:dataset}
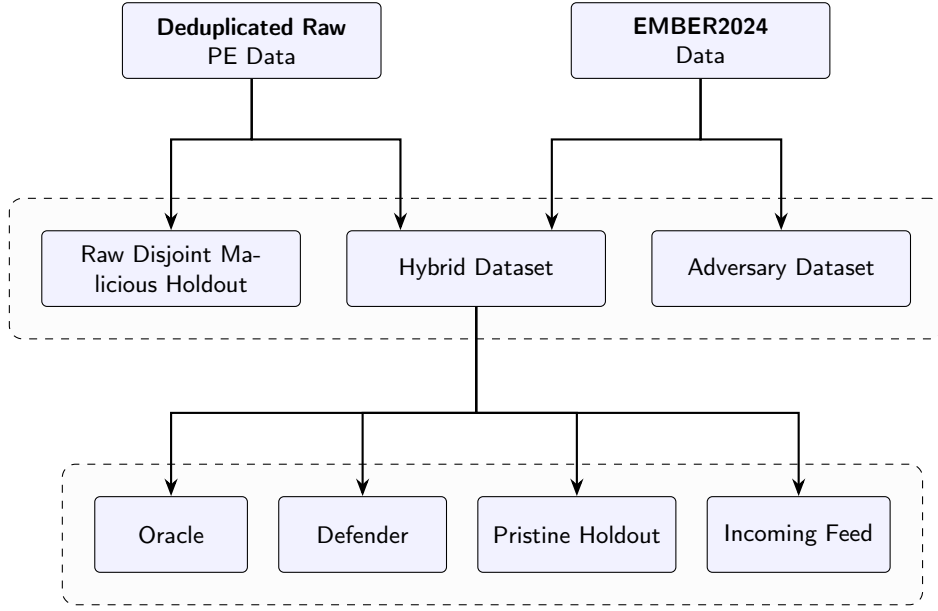
\begin{figure}[t]
\centering
\begin{tikzpicture}[
    font=\sffamily\small,
    node distance=1.2cm and 0.6cm,
    data/.style={rectangle, draw, fill=blue!5, text width=3.2cm, align=center, minimum height=1cm, rounded corners=2pt},
    proc/.style={rectangle, draw, fill=orange!10, text width=4cm, align=center, minimum height=1cm, thick},
    arrow/.style={thick, ->, >=Stealth},
    group/.style={draw, dashed, inner sep=12pt, fill=gray!2, rounded corners=5pt}
]
    \node (rawpe) [data] {\textbf{Deduplicated Raw}\\PE Data};
    \node (rawember) [data, right=2.5cm of rawpe] {\textbf{EMBER2024}\\Data};
    \node (hybrid) [data, below=2cm of $(rawpe.south)!0.5!(rawember.south)$] {Hybrid Dataset};
    \node (holdout) [data, left=of hybrid] {Raw Disjoint Malicious Holdout};
    \node (adv) [data, right=of hybrid] {Adversary Dataset};
    \draw [arrow] (rawpe.south) -- ++(0,-0.8) -| ([xshift=-1cm]hybrid.north);
    \draw [arrow] (rawpe.south) -- ++(0,-0.8) -| (holdout.north);
    \draw [arrow] (rawember.south) -- ++(0,-0.8) -| (adv.north);
    \draw [arrow] (rawember.south) -- ++(0,-0.8) -| ([xshift=+1cm]hybrid.north);
    \node (p2) [data, below=2.5cm of hybrid, xshift=-1.5cm, text width=2cm] {Defender};
    \node (p1) [data, left=0.4cm of p2, text width=1.8cm] {Oracle};
    \node (p3) [data, right=0.4cm of p2, text width=2.4cm] {Pristine Holdout};
    \node (p4) [data, right=0.4cm of p3, text width=2.2cm] {Incoming Feed};
    \foreach \n in {p1, p2, p3, p4} {
        \draw [arrow] (hybrid.south) -- ++(0,-1.4) -| (\n.north);
    }
    \begin{scope}[on background layer]
        \node (setgroup) [group, fit=(hybrid) (adv) (holdout), label={[anchor=north, font=\bfseries]}] {};
        \node (splitgroup) [group, fit=(p1) (p4), label={[anchor=north, font=\bfseries]}] {};
    \end{scope}
\end{tikzpicture}
\caption{Dataset partitioning. Raw data sources are processed, deduplicated, and merged into a hybrid dataset, which is subsequently partitioned into functional splits.}
\label{fig:dataset_pipeline}
\end{figure}

We evaluate the poisoning attack using a combination of raw PE binaries \cite{datasetlester}
and the EMBER2024 \cite{ember2024} feature-based dataset. Figure~\ref{fig:dataset_pipeline}
illustrates the data partitioning and processing pipeline.

\subsection{Data Sources and Hybridization}
The raw PE dataset contains benign files from standard OS installations and malicious
samples from VirusShare, MalShare, and TheZoo \cite{datasetlester}.
We identified near-duplicate files using Trend Micro Locality Sensitive Hash (TLSH)
with a distance threshold of 30 \cite{tlsholiver,ember2024},
resulting in a balanced dataset of 104,866 files (52,433 per class). A reduced
subset of 12,500 files was used for iterative refinement, with 1,400 malicious
samples reserved as targets.

The EMBER2024 dataset provides 3.2M feature vectors. We extracted EMBER v3 features
from the raw binaries to ensure compatibility, neutralizing potential artifacts
(e.g., timestamps, checksums) by setting them to zero. After a secondary TLSH
deduplication pass, we aggregated both sources into a unified hybrid dataset.

\subsection{Data Partitioning}
The hybrid dataset was partitioned to reflect a realistic operational environment, maintaining balanced class distributions:
\begin{itemize}[leftmargin=*, label={}]
    \item \textbf{Defender (40\%):} Initial training dataset $D_T$ for the baseline model.
    \item \textbf{Incoming Feed (20\%):} External source $D_I$ serving as the ingestion channel.
    \item \textbf{Oracle (20\%):} Subset $D_O$ used by the decision function $\Gamma$.
    \item \textbf{Holdout (20\%):} Disjoint ground-truth set $D_H$ for final evaluation.
\end{itemize}
This partitioning enables the Defender to evaluate incoming telemetry using an
ensemble, deciding whether to ingest samples or flag them for manual analysis.

\section{Experimental Evaluation}
\label{sec:experiments}
We evaluate the poisoning framework in a knowledge-asymmetry setup,
where the adversary uses public data and a proxy model to generate
adversarial binaries designed to evade pre-ingestion filters and
poison the primary detection model.

\subsection{Model Configurations}
\label{sec:models}
We simulate a realistic environment using the following components,
trained with 5-fold cross-validation:
\begin{itemize}
    \item \textbf{Defender Baseline ($M_D$)}: LightGBM model trained on
    the defender partition $D_T$, prioritizing a low FPR of $0.5\%$.
    \item \textbf{Attacker Proxy ($M_A$)}: Random Forest classifier trained
    on the public EMBER2024 partition $D_A$, approximating the defender's
    boundaries without access to proprietary data.
    \item \textbf{Homogeneous Ensemble ($\Gamma$)}: Pre-ingestion filter
    comprising three LightGBM models trained on byte distribution, content
    (strings/libraries), and structural features of the oracle partition
    $D_O$. Ingestion requires consensus across all models.
\end{itemize}

Table~\ref{tab:model_performance} summarizes component performance against the holdout set $D_H$.

\begin{table}[h]
\centering
\caption{Performance metrics of model components on the holdout dataset $D_H$.}
\label{tab:model_performance}
\resizebox{\columnwidth}{!}{%
\begin{tabular}{lccccc}
\hline
\textbf{Model} & \textbf{Threshold} & \textbf{Accuracy} & \textbf{Recall} & \textbf{FPR} & \textbf{ROC AUC} \\ \hline
Baseline Defender ($M_D$) & 0.5970 & 0.9576 & 0.9472 & 0.0320 & 0.9922 \\
Attacker Proxy ($M_A$) & 0.6321 & 0.8762 & 0.7675 & 0.0152 & 0.9724 \\
Distributional Model & 0.2544 & 0.8840 & 0.9444 & 0.1764 & 0.9641 \\
Content Model & 0.5076 & 0.9080 & 0.8920 & 0.0760 & 0.9662 \\
Structural Model & 0.5462 & 0.9426 & 0.9332 & 0.0480 & 0.9883 \\ \hline
\end{tabular}%
}
\end{table}

\subsection{Adversarial Generation and Ingestion}
We utilize the \texttt{secml\_malware} framework to generate adversarial
binaries. While whitebox gradient-based attacks (e.g., MalConv) yielded
negligible evasion rates ($0.3\%$) due to the feature space disconnect,
blackbox heuristic attacks using the GAMMA strategy \cite{gammademetrio}
proved effective. We extended GAMMA to include functionality-preserving
Import Address Table (IAT) and section injections sour\-ced from benign binaries.

Table~\ref{tab:evasion_table} summarizes the effectiveness of these
manipulations. High-intensity attacks (8 IAT/8 Section injections)
achieved a $50.8\%$ evasion rate against the proxy $M_A$.

\vspace{-0.5cm}

\begin{table*}[h]
\centering
\caption{Adversarial evasion performance and confidence impact on the attacker's 
proxy model $M_A$. \textit{IAT} and \textit{Sec.} denote the number of Import Address Table and 
Section injections, respectively. \textit{Evasion \%} denotes the fraction of 
malicious files flipped to benign ($1 \to 0$); \textit{Det. Benign} is the 
total percentage of binaries detected as benign; \textit{Drop (Flipped/Stayed)} 
represents the mean confidence drop for samples that flipped or 
remained benign, respectively.}
\label{tab:evasion_table}

\resizebox{\textwidth}{!}{%
\begin{tabular}{c c c c c c c}
\hline
\textbf{IAT} & \textbf{Sec.} & \textbf{Evasion \%} & \textbf{Det. Benign} & \textbf{Drop (Flipped)} & \textbf{Drop (Stayed)} & \textbf{Filesize Increase (KB)} \\ \hline
2 & 0 & 2.9\% & 32.6\% & 0.0460  $\pm$ 0.0557 &  0.0004 $\pm$ 0.0301 & 74.92 $\pm$ 1267.58 \\
0 & 2 & 39.0\% & 57.9\% & 0.2076 $\pm$ 0.0974 & 0.0641 $\pm$ 0.0911 & 391.64 $\pm$ 1271.37 \\
2 & 2 & 34.5\% & 54.7\% & 0.1977 $\pm$ 0.0967 & 0.0541 $\pm$ 0.0872 & 312.33 $\pm$ 1262.47 \\
5 & 5 & 47.4\% & 63.5\% & 0.2261 $\pm$ 0.0943 & 0.0799 $\pm$ 0.0939 & 632.25 $\pm$ 976.89 \\
8 & 8 & 50.8\% & 65.4\% & 0.2226 $\pm$ 0.0913 & 0.0772 $\pm$ 0.1006 & 758.49 $\pm$ 988.97 \\ \hline
\end{tabular}%
}
\end{table*}

\subsection{Poisoning Impact and Filtering Efficiency}
The poisoning attack targets the ingestion pipeline by injecting adversarial
samples $D_P$ into the incoming feed $D_I$. We evaluate an
\textit{incorrect labeling strategy} (malicious samples labeled as benign)
against a \textit{correct labeling} baseline. As shown in Table~\ref{tab:poison_naive},
subtle manipulations (2 IAT injections) yielded the most significant
recall degradation, suggesting that minimal perturbations lie closer to
the baseline decision boundary. Under a correct labeling strategy (data not shown),
performance remained stable, confirming that the observed degradation is due
to label noise.

\begin{table}[h]
\centering
\caption{Defender performance under the incorrect labeling oracle strategy.}
\label{tab:poison_naive}
\resizebox{\columnwidth}{!}{%
\begin{tabular}{c c c c c c c c}
\hline
\textbf{IAT} & \textbf{Sec.} & \textbf{Poison \%} & \textbf{Accuracy} & \textbf{Recall} & \textbf{FPR} & \textbf{ROC AUC} & \textbf{Threshold} \\ \hline
2 & 0 & 2.85\% & 0.9256 & 0.8603 & 0.0092 & 0.9918 & 0.7943 \\
0 & 2 & 5.81\% & 0.9568 & 0.9344 & 0.0208 & 0.9927 & 0.6737 \\
2 & 2 & 5.24\% & 0.9568 & 0.9348 & 0.0212 & 0.9927 & 0.6778 \\
5 & 5 & 6.68\% & 0.9552 & 0.9296 & 0.0192 & 0.9924 & 0.7012 \\
8 & 8 & 6.73\% & 0.9556 & 0.9312 & 0.0200 & 0.9922 & 0.6995 \\ \hline
\end{tabular}%
}
\end{table}

\begin{figure}[t]
    \centering
    \includegraphics[width=0.8\linewidth]{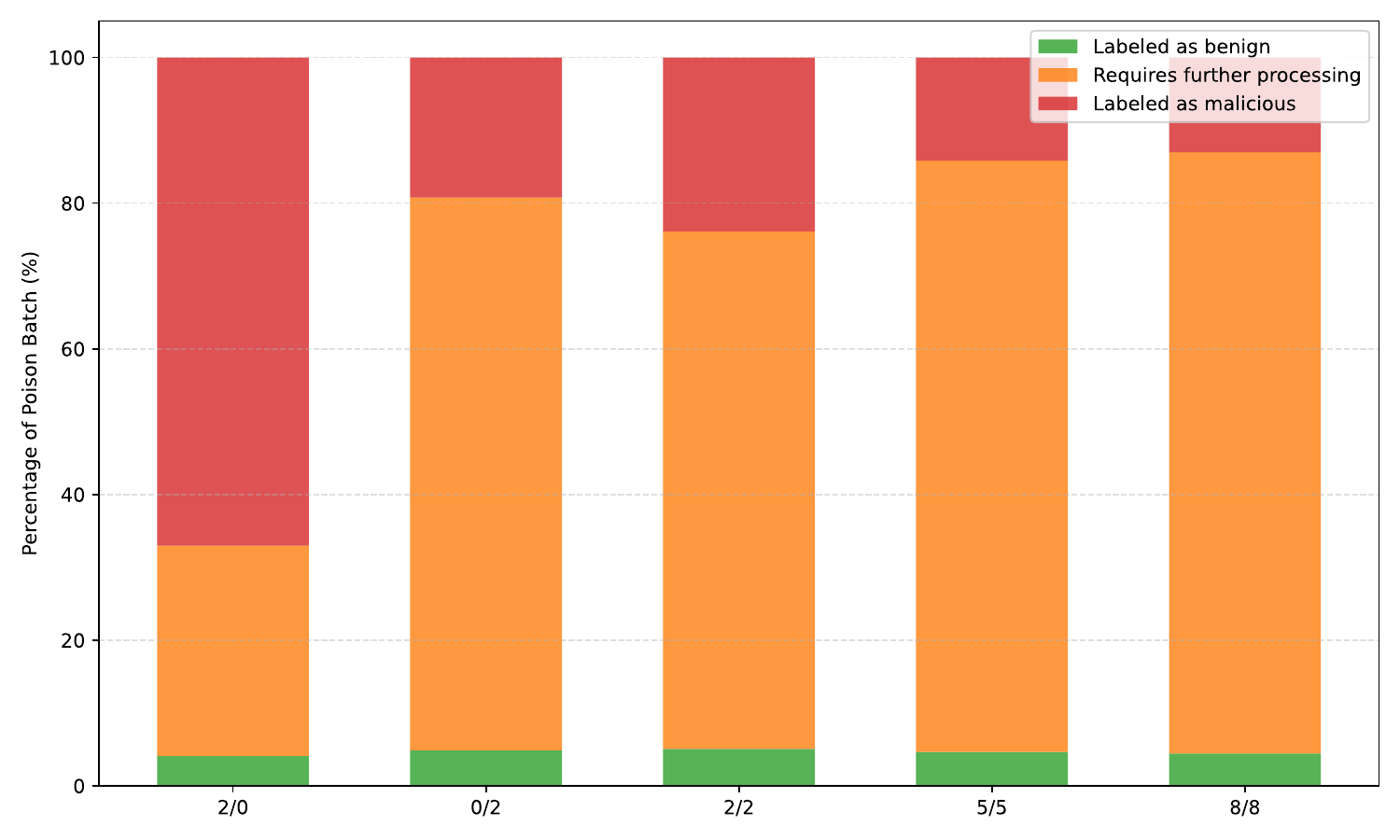}
    \caption{Ensemble agreement for poisoned samples.}
    \label{fig:ensemble_agreement}
\end{figure}

The homogeneous ensemble $\Gamma$ successfully mitigated poisoning risks,
filtering $95.6\%$ of high-intensity samples (Table~\ref{tab:filtering_efficiency})
while retaining $84.5\%$ of clean data. Figure~\ref{fig:ensemble_agreement}
shows that while IAT injections effectively poison the model, they remain
highly detectable by structural oracles, forcing a trade-off between stealth
and efficacy.

\vspace{-0.5cm}

\begin{table}[h]
\centering
\small
\caption{Filtering efficiency of individual and combined oracles (Intensity: 8 Sec., 8 IAT).}
\label{tab:filtering_efficiency}
\begin{tabular}{lc}
\hline
\textbf{Model} & \textbf{Filtered Rate} \\ \hline
Distributional Model & 25.0\% \\
Content Model & 18.4\% \\
Structural Model & 94.2\% \\ \hline
\textbf{Combined (Any)} & \textbf{95.6\%} \\ \hline
\end{tabular}
\end{table}

As shown in Figure~\ref{fig:ensemble_agreement},
IAT injections are effective for model poisoning but highly detectable by 
the homogeneous ensemble. Conversely, the injection of benign sections 
effectively shifts the decision boundary and more frequently necessitates 
secondary verification. This implies that an adversary identifies an 
optimal balance, performing minimal structural modifications to 
maintain stealth while maximizing the degradation of defender confidence.

\section{Limitations and Future Work}
\label{sec:limitations}
This preliminary evaluation utilizes a reduced subset of 12,500 samples;
future work will scale to the complete hybrid dataset to ensure generalizability
across larger and more diverse malware samples.
While we assume similarity in feature distributions between the defender
and adversary, investigating the transferability of adversarial samples
across distinct feature representations is essential for a full black-box
setting. Furthermore, our study is limited to single-step poisoning events,
whereas real-world pipelines operate continuously. Persistent poisoning over
multiple intervals and the use of advanced architectures, such as deep
ensemble methods or graph-based representations of PE structures, remain
key areas for future investigation.
\section{Conclusion}
\label{sec:conclusion}
This work demonstrated a realistic gray-box poisoning threat model for continuous
malware ingestion pipelines. By leveraging the \texttt{secml\_malware} framework,
we showed that IAT-based perturbations enable compact poisoning samples that significantly
degrade detection recall, while more intensive modifications are easily detected by
pre-ingestion filters. These results highlight a critical trade-off between adversarial
stealth and poisoning efficacy. Our proposed homogeneous ensemble filter successfully
mitigated 95.6\% of poisoning attempts, emphasizing the importance of multi-faceted
validation in production pipelines. Future research will explore these methodologies
in full black-box and iterative poisoning scenarios.

\section*{Acknowledgements}
This work was supported by the Grant Agency of the Czech Technical University in Prague, grant No. SGS26/187/OHK3/3T/18 funded by the MEYS of the Czech Republic.

\bibliographystyle{splncs04}
\bibliography{example}


\end{document}